\documentclass[sigconf,authorversion,nonacm]{acmart}

\makeatletter                   
\def\mdseries@tt{m}             
\makeatother                    

\renewcommand\footnotetextcopyrightpermission[1]{} 
\AtBeginDocument{%
  \providecommand\BibTeX{{%
    \normalfont B\kern-0.5em{\scshape i\kern-0.25em b}\kern-0.8em\TeX}}}

\acmConference[AISec '20]{AISec '20: 13th ACM Workshop on Artificial Intelligence and Security}{November 13, 2020}{Orlando, FL}

\begin{document}
\fancyhead{}

\title{Flow-based detection and proxy-based evasion  \\ of encrypted malware C2 traffic}

\author{Carlos Novo}
\affiliation{
  \institution{University of Porto and INESC TEC}
  \city{Porto}
  \country{Portugal}}
  \email{carlos.novo@fe.up.pt}

\author{Ricardo Morla}
\affiliation{
  \institution{University of Porto and INESC TEC}
  \city{Porto}
  \country{Portugal}}
  \email{ricardo.morla@fe.up.pt}

\begin{abstract}
State of the art deep learning techniques are known to be vulnerable to evasion attacks where an adversarial sample is generated from a malign sample and misclassified as benign. Detection of encrypted malware command and control traffic based on TCP/IP flow features can be framed as a learning task and is thus vulnerable to evasion attacks. 
However, unlike e.g. in image processing where generated adversarial samples can be directly mapped to images, going from flow features to actual TCP/IP packets requires crafting the sequence of packets, with no established approach for such crafting and a limitation on the set of modifiable features that such crafting allows.
In this paper we discuss learning and evasion consequences of the gap between generated and crafted adversarial samples. We exemplify with a deep neural network detector trained on a public C2 traffic dataset,  white-box adversarial learning, and a proxy-based approach for crafting longer flows. 
Our results show 1) the high evasion rate obtained by using generated adversarial samples on the detector can be significantly reduced when using crafted adversarial samples;  2) robustness against adversarial samples by model hardening varies according to the crafting approach and corresponding set of modifiable features that the attack allows for; 3) incrementally training hardened models with adversarial samples can produce a level playing field where no detector is best against all attacks and no attack is best against all detectors, in a given set of attacks and detectors.
To the best of our knowledge this is the first time that level playing field feature set- and iteration-hardening are analyzed in encrypted C2 malware traffic detection.

\end{abstract}

\ccsdesc[500]{Security and privacy~Intrusion/anomaly detection and malware mitigation}
\ccsdesc[500]{Computing methodologies~Machine learning}

\keywords{Malware command and control, intrusion detection, adversarial learning}

\maketitle

\section{Introduction}

Detecting encrypted malware command and control traffic using machine learning and traffic characteristics is important -- especially for zero-day attacks or when attackers frequently change black-listed items such as IP addresses or server-side certificates. However, machine learning algorithms are known to be vulnerable to adversarial attacks and it is only natural that attackers would explore this vulnerability and modify the behavior of C2 traffic between victim and C2 server to make detection harder. But malware development is costly. Even with a large set of open source malware frameworks to choose from, modifying the behavior of complex code in order to achieve specific adversarial traffic can be a task with daunting impact on the profit of most operations. Applying traffic proxies or addons to the source code that do not modify the behavior of the malware but can modify traffic features can be a more attractive, lower cost solution. Deploying modified malware on the C2 server and even on victims seems to be a comparably easier problem with mechanisms available for overwriting old payloads with new ones on the victims. 

\begin{figure}[h]
    \centering
    \includegraphics[width=0.48\textwidth]{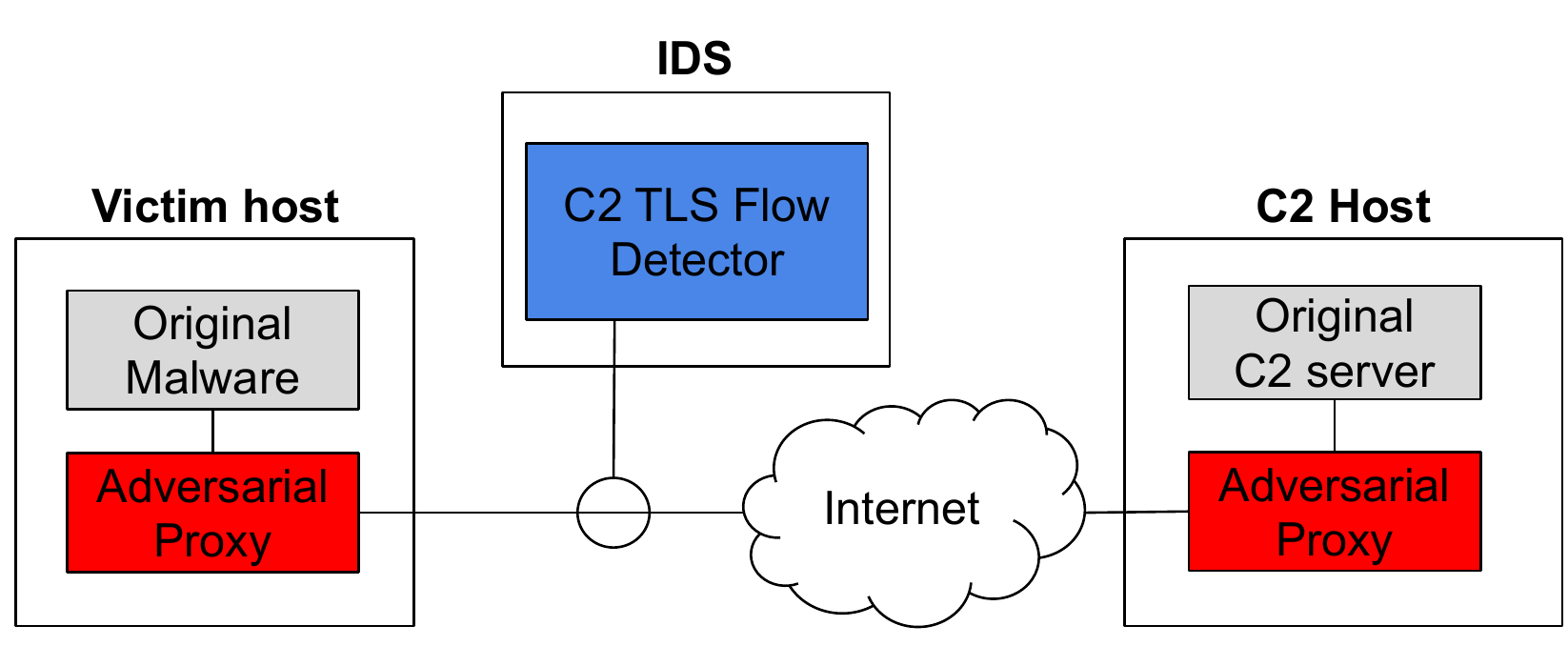}
    \caption{Proxy-based modification of TLS C2 flows. }
    \label{fig:proxy-based-attack}
\end{figure}

We take these constraints as the motivation for this paper. Figure \ref{fig:proxy-based-attack} illustrates an attack and defense architecture compatible with these constraints. The adversarial proxies or addons in the C2 server and the victim monitor the traffic between the original malware and C2 server, and when necessary -- for example just before the end of the TCP connection -- delay or add packets thus modifying traffic characteristics that do not change the behavior of the original malware and hopefully can break the detector. The detector, sitting in one end of the Intrusion Detection System, complements existing rule-based modules in the IDS by detecting yet-to-blacklist C2 traffic with a machine learning model. 

While attackers are said to be always more motivated than defenders, our approach in this paper is to try and set a level playing field between attacker and defender. Both attacker and defender start from a common public dataset and we set out to understand what happens when each develops their own strategy and these strategies collide. In particular we try to look at attacker and defender performance for different setups of the attacker and defender -- namely what happens if a set of features is used for attacking and another set of features for hardening the model, and if the attacker goes through more or less hardening and attacking steps than the defender. Although we do not explicitly use proxy models for the attacker to generate adversarial samples, recent work~\cite{demontis2019adversarial} points to adversarial attacks transferring from one model to another -- in our case we assume the deep learning model structure and the original training data is the same for attacker and defender.

The remaining of this section lays out assumptions for the underlying traffic, datasets, and limitations to how much the malware can be changed. Section \ref{sec:mta-dataset} describes the dataset that we use in the paper. Section \ref{sec:detection-and-evasion} describes the deep learning model used for detection, the method used for generating adversarial samples, an example crafting technique to increase the duration of the flow in actual packet traces, and performance results for the defender under attacks with different crafting and feature set limitations. Sections \ref{sec:hardening} and \ref{sec:loop} explore the level playing field clash between attacker and defender, with different possibilities for attacker and defender strategies in hardening and iterating through attack-hardening steps to improve detector resilience and adversarial sample impact. We conclude with a review of the state of the art in section \ref{sec:related-work} and final notes in section \ref{sec:conclusions}.

\subsection{Assumptions}

\subsubsection{TLS traffic}
Deep packet inspection and signature-based techniques can be used to detect plaintext C2 traffic, but are of limited usefulness for encrypted traffic. Certificate- and cipher suite-based techniques have been proposed to detect encrypted malware~\cite{anderson_deciphering_2018}, but are not robust against C2 servers changing certificates and ciphers. Here we assume some other approaches will be used for detecting plaintext C2 traffic and filtering black-listed server IPs and certificates, but that frequently changing certificates will require a detector that focuses on the characteristics of TLS traffic. 

\subsubsection{Models built on publicly available datasets and tools}
We limit our analysis to a dataset that is publicly available to both the defender and the attacker. This allows the attacker to build a proxy model and develop their attack from the proxy model if the actual model is not available. We consider both cases where the defender model is available to the attacker (e.g. if it was stolen) and where a proxy model is built from the publicly available data that the defender used. We do not consider the case where data from the same malware is captured from other environments and used by the attacker.

\subsubsection{Limitations to how much the malware can be changed}
We assume the malware developer is unwilling or unable to change the code for the main functionality of the malware. In this case there are two options for changing C2 traffic: 1) a proxy-like approach that works in the non-TLS part of the communications, e.g. increasing the duration of the flow by hanging on to the FIN/ACK packets at the end of the flow long enough to reach the larger flow duration adversarial target; 2) adding adversarial code at the end of the original malware code that does not change the original code but that sends additional TLS records which the C2 server receives but ignores, adding to the byte and packet count but also possibly changing duration. Figure \ref{fig:proxy-based-attack} illustrates the proxy-like attack. This assumption is strong and we expect to address weaker assumptions in future work namely using open source command and control frameworks like Metasploit that we can rewrite and use to more closely resemble the target flow features.

\section{Malware Traffic Analysis Dataset}
\label{sec:mta-dataset}

\subsection{Scraping}

We use data from \url{https://malware-traffic-analysis.net} (MTA), which contains recent, vast, 
and detailed content related to common malware families. The malware is ran in sandboxed environment and specific malware infections are described, many of them providing
.pcap files containing captured network traffic associated to the infection and to the C2 communication. After scraping the website, 508 files were obtained, corresponding to captures from 2016 to 2019. Table \ref{tab:tls_per_year} describes the years and statistics of the data we used. Although each .pcap file has the date on which it was added to the MTA website, in a real network we could have traffic from older and newer malware families coexisting and because of that and of the small number of flows in our dataset we do not further distinguish malware based on year or malware family. 

\begin{table}[ht]
  \centering
\begin{tabular}{|c|rrr||r@{.}l|}
	\hline
    \textbf{\em year}
	& \# files fetched
	& \# TCP flows
	& \# TLS 
	& \multicolumn{2}{c|}{ \% TLS} \\ \hline \hline
2016   & 109  & 1213     & 58    &  4&782\%    \\
2017   & 141  & 3168     & 885   & 27&936\%  \\
2018   & 147  & 24854    & 11590 & 46&632\%   \\ 
2019   & 111  & 39286    & 8214  & 20&908\% \\ \hline
total  & 508  & 68521    & 20747 & 30&278\% \\ \hline 
\end{tabular}
  \caption{Per year statistics of the MTA files we fetched.}
  \label{tab:tls_per_year}
\end{table}

\subsection{Labeling individual C2 flows}

We used the SSL Blacklist project\footnote{\url{https://sslbl.abuse.ch}} list of SSL/TLS certificates employed by botnet C2 servers, and configured it on Suricata\footnote{\url{https://suricata-ids.org}} with rules to identify malicious TLS flows. Table \ref{tab:mal-families} shows how the 7672 malicious TLS flows are distributed per black-listed malware family certificates. We considered the remaining 13075 TLS flows as normal.

\begin{table}[h]
    \centering
    \begin{tabular}{c|c}
         \textbf{Family}    & \textbf{Flow count} \\
         TrickBot           & 4984 \\
         PandaZeuS          & 1610 \\
         Gozi               &  436 \\
         IcedID             &  374 \\
         Dridex             &  131 \\
         AKBuilder          &   56 \\
         IcedId             &   56 \\
         Others {\footnotesize($<50$ flows)}  &   25  \vspace{3mm}
    \end{tabular}
    \caption{Number of C2 TLS flows per malware family}
    \label{tab:mal-families}
\end{table}

\subsection{Class imbalance and training-test split}

We use all of the malicious TLS flows in the MTA dataset and randomly take the same number of normal TLS flows to create a balanced dataset. From this, we randomly take 20\% for testing. If capturing data on a real network, the imbalance between normal and malicious would be much larger and the test set should reflect this imbalance. We plan to consider class imbalance in future work where we setup real network data capture.

\subsection{Obtaining flow features}

We use TStat~\cite{tstat} to extract 86 numerical features from traffic flows, including the total number of packets observed from the client or server and the duration of the flow. We ignore incomplete flows as defined by Tstat. Table \ref{tab:tstat-features} lists the specific features\footnote{\url{http://tstat.polito.it/measure.shtml\#log\_tcp\_complete}} we used. These are mostly numerical. We normalize features between 0 and 1 by dividing by the maximum value for the feature in the training set and square rooting: $f^i_n = \sqrt{f^i/f^i_{max}}$. Features in the test data whose value is larger than $f^i_{max}$ are set to 1.

\begin{table}[h]
    \centering
    \begin{tabular}{c|c}
         \hline
         \hline
        \textbf{Core Set}          & \textit{Features 3-14, 17-28, 31-37} \\
        \hline
         3-6 ; 17-20      & Total, RST, ACK, pure ACK \\
         &packet counts \\
         7 ; 21    & Unique bytes \\
         8-9 ; 22-23     & Data segment and byte counts \\
         10-11 ; 24-25     & Retransmitted data segment \\
         & and byte counts \\
         12 ; 26    & Out of sequence segment counts \\
         13-14 ; 27-28 & SYN and FIN packet counts \\
         31 & Flow Duration \\
         32 ; 33 & Rel. time of first payload \\
         34 ; 35 & Rel. time of last payload \\
         36 ; 37 & Relative time of first ACK \\
         \hline
         \hline
                  \textbf{TCP End to End} & \textit{Features 45-58} \\
        \hline 
        45-48 ; 52-55 & Av., min., max., st.dev. RTT\\
        49 ; 56 & RTT observation counts \\
        45-51 ; 57-58 & Maximum and minimum TTL  \\
         \hline
         \hline
         \textbf{TCP Options} &  \textit{Features 65-79,83,85,90-104,106-109}\\
         \hline
         65-66 & Window scale, timestamp \\
         &options sent (C2S)\\
         67 ; 90 & Scaling values negotiated \\
         68-69 ; 91-92 & SACK option set, SACK counts \\
         70-72 ; 93-95 & MSS declared, max. and min. MSS values \\
         73-74 ; 96-97 & Max. and min. receiver \\
         &window announced \\
         75 ; 98 & zero receiver window counts \\
         76-78 ; 99-101 & Max., min., initial congestion window \\
         79 ; 102 & Retransmitted by timeout counts/RTO \\
         103 & Retransmitted by 3 dup-ack \\
         &counts/FR (S2C) \\
         104 & Packet reordering counts (S2C) \\
         106 & Unknown packet counts (S2C) \\
         107 & Probe the receiver window \\
         &counts (S2C) \\
         85 ; 108 & Unnecessary retransmission \\
         &counts by RTO \\
         109 & Unnecessary retransmission \\
         & counts by FR (S2C) \\
        \hline 
        \hline
         TCP Layer 7 &  \textit{Features 114,115,120-122} \\
        \hline
         114 ; 115 & Push-separated message counts \\
         120 & Client TLS session ID reuse \\
         121 ; 122 & Rel. time of last packet \\
         &before first TLS App. record \\
         \hline
         \hline
    \end{tabular}

    \caption{TStat features used to characterize flows. We use \textit{C2S ; S2C} feature range notation to represent client-to-server and server-to-client feature indices.}
    \label{tab:tstat-features}
\end{table}

\begin{figure*}[hbt]
  \begin{center}
    \leavevmode
    \includegraphics[width=\textwidth]{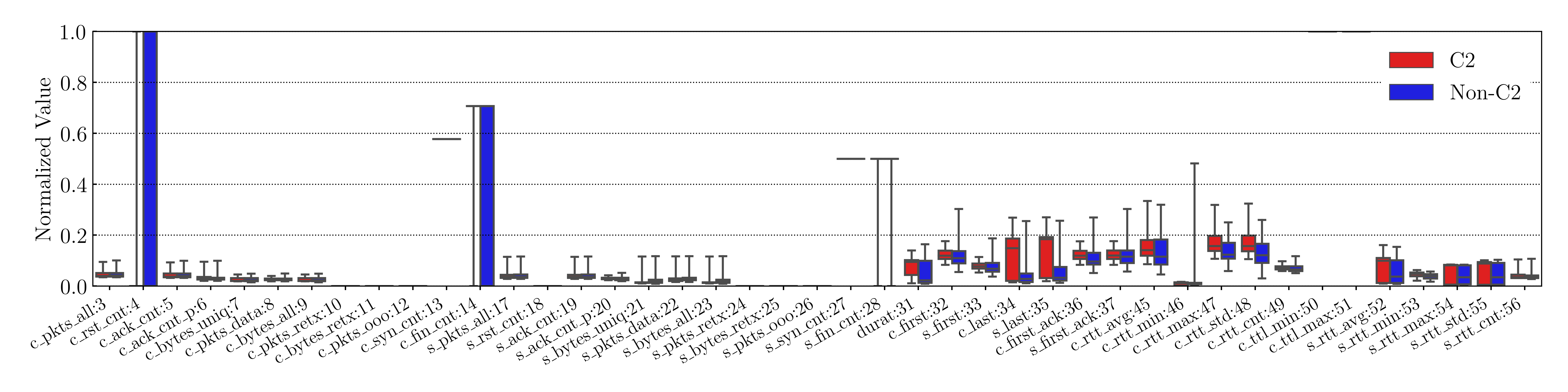}
    \includegraphics[width=\textwidth]{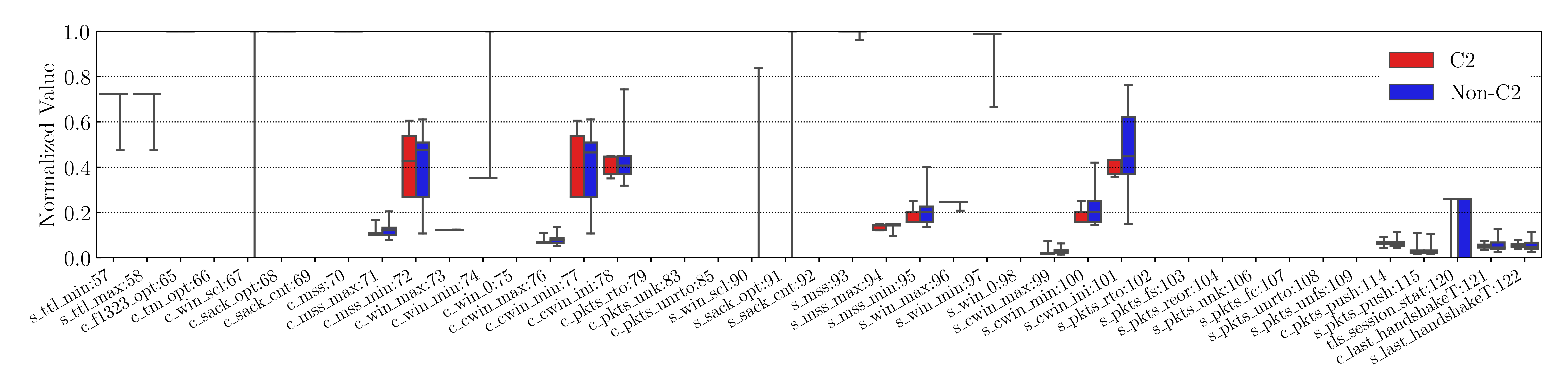}
    \caption{Boxplot of the 86 normalized flow features in the training set, for malicious (C2) and normal (Non-C2) TLS flows.}
    \label{fig:nf_boxplot}
  \end{center}
\end{figure*}

Figure \ref{fig:nf_boxplot} shows the boxtplot of the normalized feature values in our training dataset for malicious and benign TLS flows. Approximately half of the features have low entropy and the limited range of values they take can be observed on the boxplots. The remaining features have a more dynamic range and higher entropy. We can also observe that there are some differences between the C2 and non-C2 ranges of values for some features, which could help detect malicious TLS flows.

\begin{figure}[hbt]
  \begin{center}
    \leavevmode
    \includegraphics[width=0.49\textwidth]{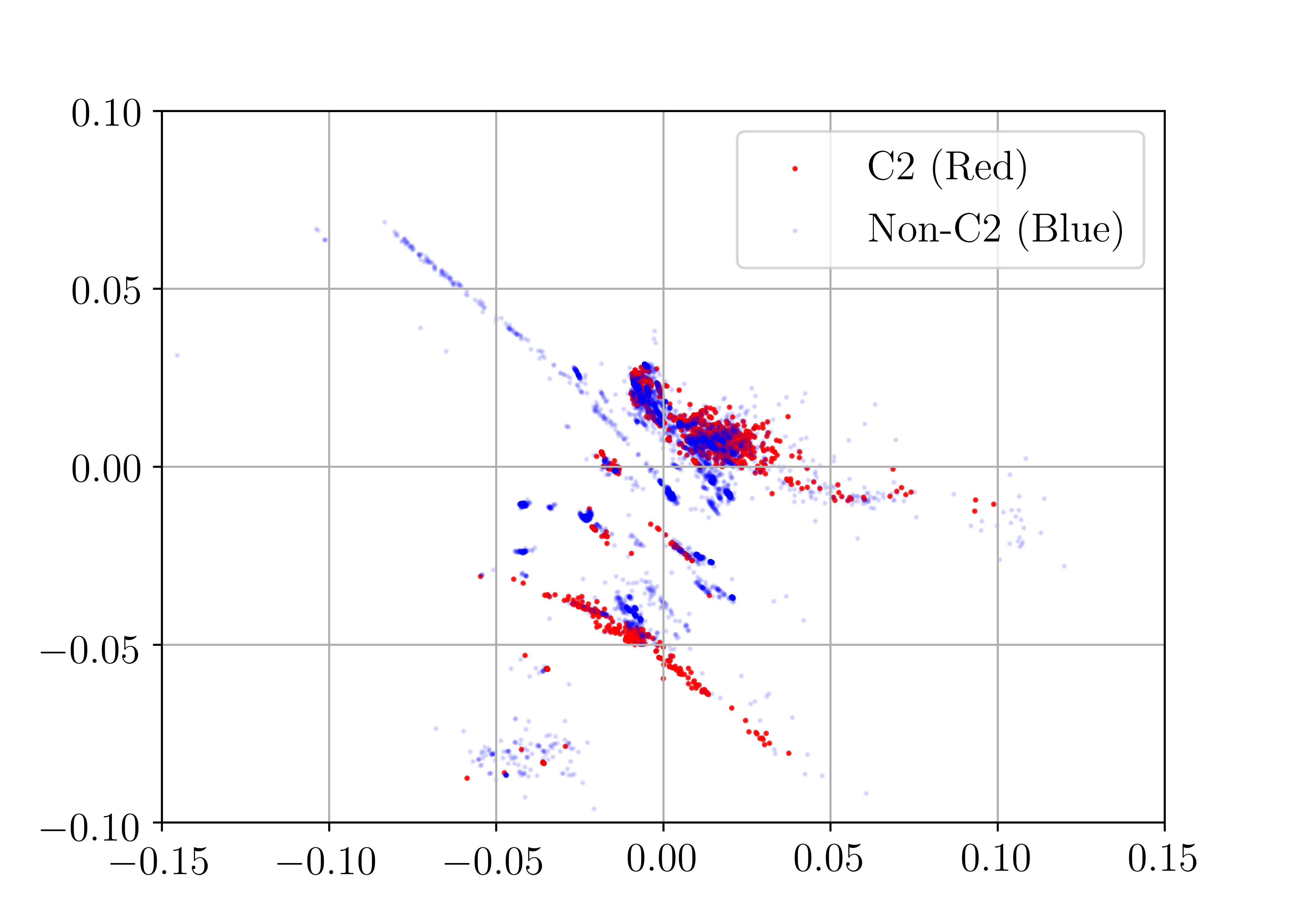}
    \caption{Visualization of the MTA training dataset.}
    \label{fig:2d_mal_ben}
  \end{center}
\end{figure}

Figure \ref{fig:2d_mal_ben} shows malicious and normal TLS flows in our training dataset and could help understand the difference between the two classes. For the visualization we used an autoencoder with 2-neuron code, 512-neuron hidden layer, 86 inputs, and 86-neuron output layer, trained with the normalized flow features in the training data. We colored the flows according to the groundtruth. Several groups of flows are visible, with malicious/benign overlapping in some of the groups. Note that this does not necessarily apply to the detector, which will use more dimensions for classification and will possibly be able to distinguish the 2D-overlapping groups.

\section{Detection and Evasion}
\label{sec:detection-and-evasion}

\subsection{DNN detector}

The detector used throughout the paper is an 86 input, 3-layer fully connected neural network with 2048/1024/512 ReLU activated neurons with 0.2 Dropout layer, and an 2-neuron Softmax activation output layer. The training uses the Adam optimizer with categorical crossentropy loss. The model has +2.8M parameters. We assume this is a generic neural network structure that can be plausibly used by anyone building a neural network-based detector. 

\subsection{FGSM whitebox attack with domain-specific restrictions}

With the Fast Gradient Sign Method~\cite{goodfellow_explaining_2015} we take a malicious C2 flow, compute the gradient of the detector's loss with regard to the input flow, and take a fixed length step $\epsilon$ for each feature, in the direction of the gradient, 
$x^* = x + \epsilon \, sign( \nabla_{x} (\theta, x, y)).$
We use CleverHans~\cite{papernot2018cleverhans} to implement this attack. 

Table \ref{tab:tstat-features-attack} defines three subsets of \textit{attack features} that can be modified given our assumptions about what the attacker can change in the C2 flow. Set 1 is for an attack that only changes the duration of the attack; set 2 is for an attack that in addition to duration changes the total number of bytes and packets; set 3 is the Tstat Core set without 4 features that measure timings at the beginning of the flow, which falls out of our assumptions for the attacker.

For each adversarial sample generated by the iterative FGSM method we take two additional steps: 1) only modify the values of the original samples for the \textit{attack features}; 2) only use the values of the \textit{attack features} if they are larger than the their corresponding values in the original sample, which is valid for counter features like bytes, packets, and flow duration and because the practical attack cannot decrease packets, byte counts, and duration of the flows. 

\begin{table}[h]
    \centering
    \begin{tabular}{c|c}
         \textbf{Attacker feature set}    & \textbf{Feature indices} \\
         Set 1 -- Duration            & 31 \\
         Set 2 -- Basic  & 31, 3/17 (all packets), 9/23 (all bytes) \\
         Set 3 -- Core &  Core set features \\
         & except 32-33 and 36-37\\
    \end{tabular}
    \caption{Attack features}
    \label{tab:tstat-features-attack}
\end{table}

\subsection{Proxy-based crafting of longer flows}

The proxy-based attack in this paper considers a proxy deployed by the attacker with the purpose of altering the characteristics of the C2 TLS flows. To change the total duration of the flow, the proxy withholds the final TCP FIN packets for the intended amount of time. To change the number of packets and bytes, the proxy injects additional packets with the desired number of bytes.  If the proxy is deployed in both ends (payload on the victim side and C2 server on the attacker side) then both client and server packet and byte count can be increased. In order to generate the adversarial sample the proxy captures the C2 flow packets and, once the final TCP FIN packet is received, sends the packets to Tstat and obtains the flow statistics that are input to FGSM. 

\begin{figure}[h]
    \centering
    \includegraphics[width=0.4\textwidth]{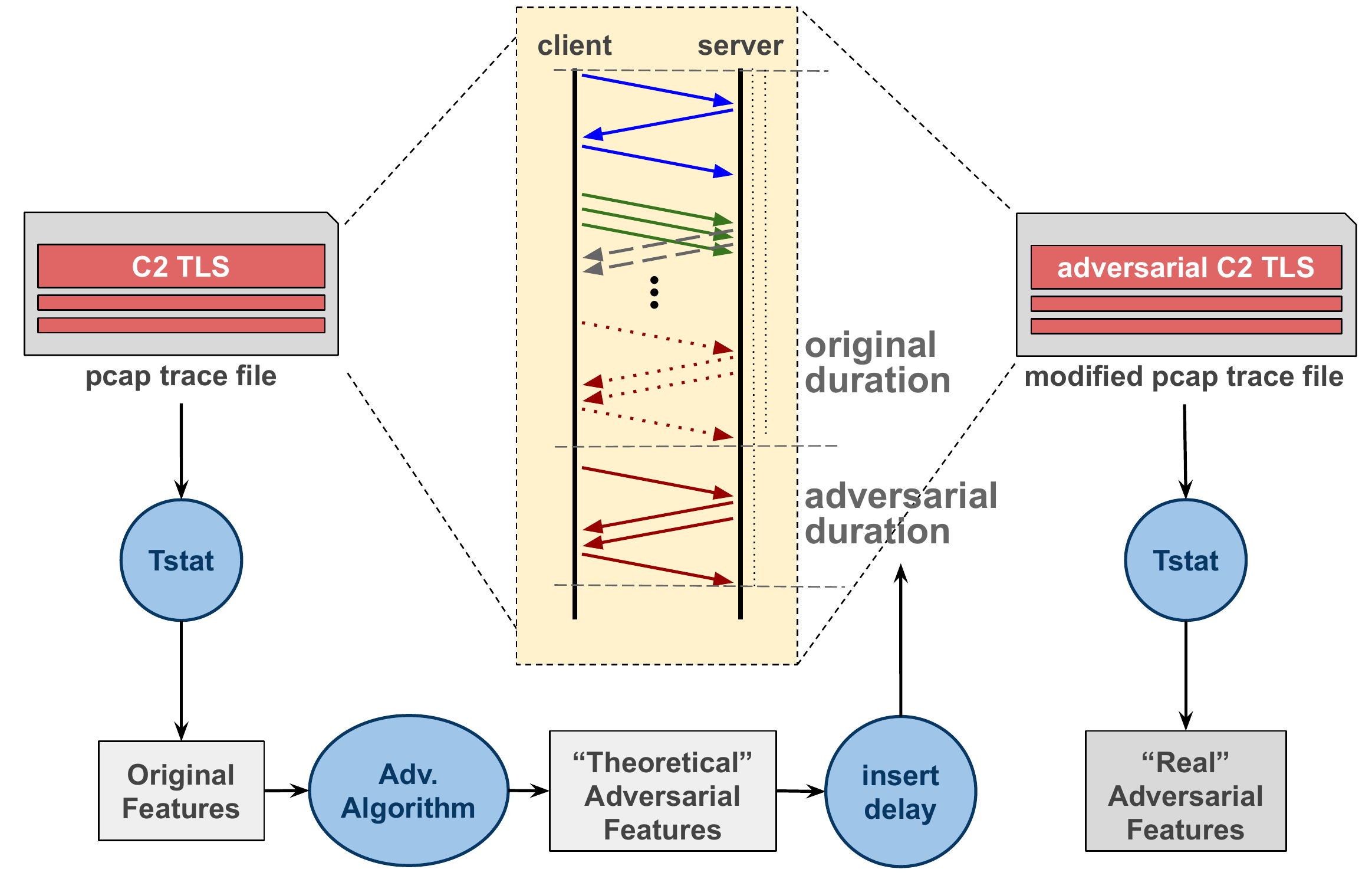}
    \caption{Obtaining adversarial examples from traffic modification by changing flow duration}
    \label{fig:traf-mod}
\end{figure}

To have a sense of real impact on all features we implemented the adversarial attack on the duration feature by modifying the MTA packet trace files as follows. Using scapy we modify the timestamp of the last four packets of the flow according to the adversarial sample, in practice causing a delay at the end of the flow that increases flow duration. The original features of the C2 flows are fed to FGSM, and the target adversarial duration feature value is de-normalized by squaring and multiplying by the maximum value. 
Figure \ref{fig:traf-mod} illustrates this process.

\subsection{Detection and Evasion Performance}

Table \ref{tab:det_ev_results} shows accuracy, precision and recall results for detection and evasion. We take the original malicious samples, change their value according to each of the attacks, and then assess performance on a data set consisting of the original benign samples and the modified malicious samples.

\begin{table}[ht]
  \centering
\begin{tabular}{|r|c|r|r|r|}
	\hline
   & &  Accuracy & Precision & 1 - Recall \\ 		
   & &   &  & (Attack \\ 		
   & &   &  & Success) \\ 		
    \hline 
    \hline
Original  & & 95.0\% & 92.7\% & 2.7\% \\
Adv. All Features & +/-  &  50.1\% & 39.5\% & 95.0\% \\
\hline
Adv. Duration & +/-  & 67.7\% & 84.3\% & 58.8\% \\
Adv. Duration & +  & 79.0\% & 89.4\% & 35.5\% \\
Crafted Duration & + & 83.1\%  & 90.2\% & 26.1\% \\ 
2x Duration & + & 93.9\% & 92.6\% & 4.9\% \\
5x Duration & + & 68.1\% & 84.6\% & 57.9\% \\
10x Duration & + & 63.9\% & 81.3\% & 66.7\% \\
20x Duration & + & 58.1\% & 73.7\% & 78.6\% \\
100x Duration & + & 51.6\% & 51.3\% & 91.9\% \\
\hline
Adv. Set 2 & +/- & 64.7\% & 82.1\% & 65.0\% \\
Adv. Set 2 & + & 67.6\% & 84.3\% & 58.9\% \\
Adv. Set 2 client & +/- & 67.2\% & 84.0\% & 59.7\% \\
Adv. Set 2 client &  + & 71.4\% & 86.5\% & 51.2\% \\
Adv. Set 2 server & +/- & 60.5\% & 77.6\% & 73.5\% \\
Adv. Set 2 server & + & 70.3\% & 85.9\% & 53.3\% \\
\hline
Adv. Set 3 & +/-  & 54.3\% & 64.2\% & 86.3\% \\
Adv. Set 3 & + &  55.6\% & 68.1\% & 83.7\% \\
Adv. Set 3 client & +/- & 58.8\% & 75.0\% & 77.0\% \\
Adv. Set 3 client & + & 64.3\% & 81.8\% & 65.7\% \\
Adv. Set 3 server & +/- & 54.0\% & 63.2\% & 86.9\% \\ 
Adv. Set 3 server & + & 54.4\% & 64.4\% & 86.1\% \\

\hline
 \end{tabular}
  \caption{Detection and evasion results. According to the attack we label the data '+/-' if the feature values can be both increased and decreased, and '+' if the feature values are only increased. 'client' and 'server' indicate whether the attack modified client or server features.}
  \label{tab:det_ev_results}
\end{table}

After training the DNN classifier on the MTA dataset we achieved a 95.0\% accuracy on the test data, with 2.7\% malicious flows and 7.3\% normal flows misclassified. Running a single iteration FGSM with $\epsilon = 0.3$ and allowing for changes in any feature increases misclassified malicious flows to 95.0\%. If we restrict the changes to the duration feature but allow decreasing flow duration (which is out of scope for our assumptions), misclassified malicious flows drop to 58.8\%. If we keep under our assumptions and only allow increasing flow duration, then misclassified malicious flows drop to 35.5\%. Although a 35.5\% success at evading the classifier seems low, it causes more than 10 times more malicious flows to be misclassified than originally, for a relatively simple attack. Finally, if we consider the changes in the pcap file described in figure \ref{fig:traf-mod}, this value further drops to 26.1\%. The attack is slightly less effective when actually implementing it in the packet traces likely because other features are modified by the shift in packet timestamps done in the attack -- e.g. the RTT features.

\begin{figure}[h]
    \centering
    \includegraphics[width=0.48\textwidth]{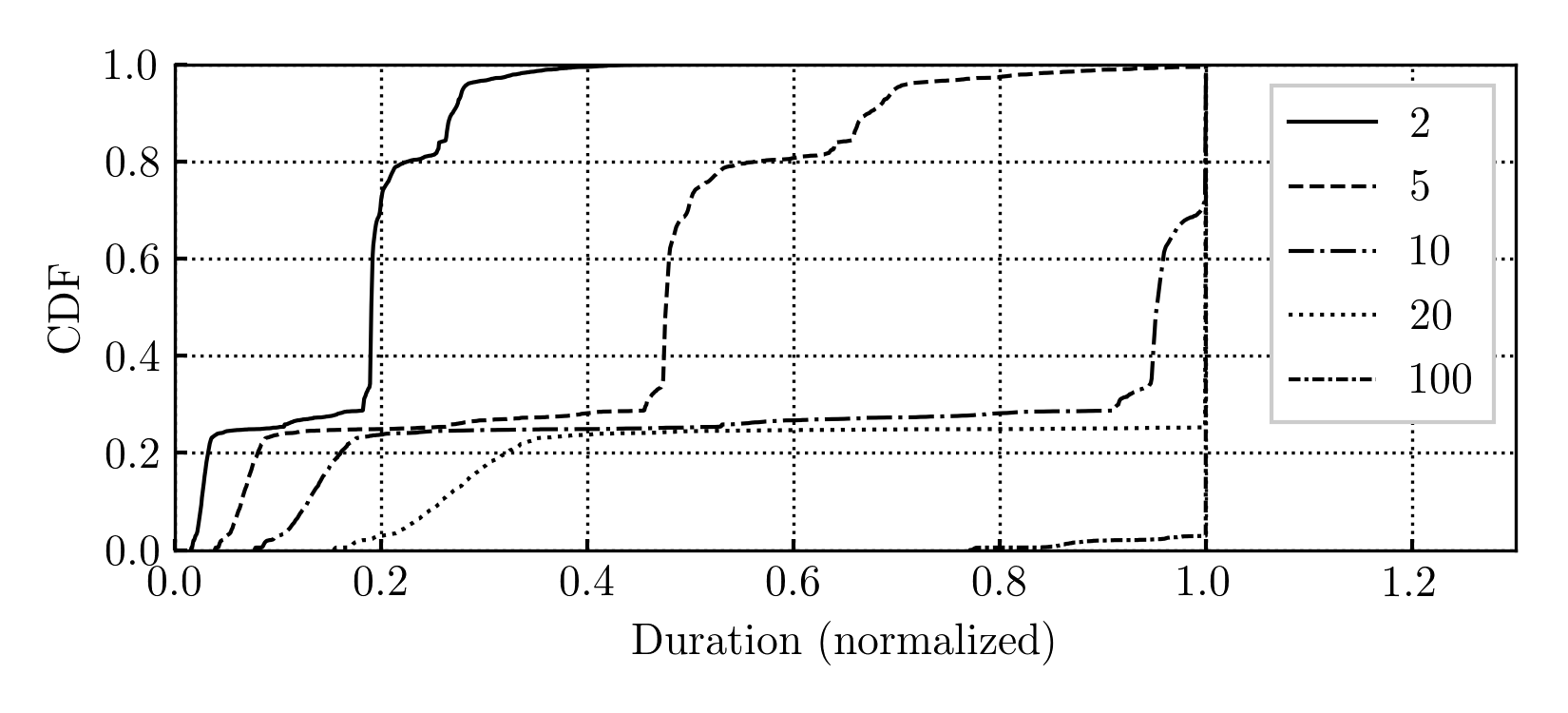}
    \caption{Increased duration CDF for simple duration multiplication attack. Legend values are the factor multiplied by the original duration.}
    \label{fig:duration_xtimes}
\end{figure}

We compared the FGSM duration attack with even simpler attacks that double -- or multiply by x -- the duration of the malicious flow and don't have to obtain flow features other than duration nor apply the FGSM method. We observe that larger percentages of malicious flows are misclassified with larger duration multiplication factor. For 100x duration we get 91.9\% misclassified malicious flows; however, in figure \ref{fig:duration_xtimes} we can see from the CDF of the flow durations for attacks x10, x20, and x100 that these flows are much larger than $episolon=0.3$, which is the maximum increase caused by our FGSM duration attacks. In particular for x100 duration is 1.0 in practically all flows. Although increasing duration does not have a direct cost, for C2 that is repetitive and frequently opens and closes TLS connections it may lower the frequency with which the malware contacts the C2 server.

Using more features than just duration -- such as those in sets 2 and 3 in table \ref{tab:tstat-features-attack} -- appears to improve the attack performance, as expected given the high level of performance we observe by using all features. Set 2 (with increase of feature value only) has 58.9\% misclassified malicious flows which is similar to the 5x duration attack; set 3 (also with increase of feature value only) has 83.7\%  misclassified malicious flows which is larger than the 20x duration attack. Notice that both set 2 and set 3 attacks only add up to +0.3 to the normalized features in the set, whereas from figure \ref{fig:duration_xtimes} most 20x flows have maximum  normalized duration (1.0). We also observe that a penalty is implied if it is only possible to modify the client- or server-side features. The notable exception is for server-side feature set 3 which appears to perform as well as the full set -- both with increasing and decreasing feature values or only with increasing feature values. This would call for setting up the proxy at the C2 server.



\section{Hardening the detector}
\label{sec:hardening}

In this section we try to understand not only to which extent hardening the detector is possible (we expect it is given prior work on adversarial learning) but especially how robust a model hardened with adversarial samples from a given attack is against adversarial samples from other attacks, considering the flow-specific limitations in the set of features in the attacks. To harden the detector we take the original data set and extend it with adversarial samples. We train the hardened model on the extended training data set and assess the performance of the hardened detector on both the original and the adversarial test data. 

\subsection{Robustness against other adversarial samples}

Table \ref{tab:hardening-results} shows hardening and attacking results for some of the feature sets in table \ref{tab:det_ev_results}. The small values in the matrix diagonal in the table shows us that all hardened models, having been trained on adversarial samples with a given feature set, are robust to a test set of adversarial samples of the same feature set. The following paragraphs provide a read-though and analysis of the non-diagonal elements of the matrix.


\begin{table*}[ht]
\small
  \centering
\begin{tabular}{|r|c||r||r|r|r|r||r|r|r|r||r|r|r|r|}
	\hline
&Index&1&2&3&4&5&6&7&8&9&10&11&12&13\\
\hline
\hline
Adv. All Features +/-&1&0.1\%&36.0\%&\textbf{36.1\%}&\textbf{84.6\%}&97.6\%&45.2\%&35.0\%&\textbf{47.8\%}&25.1\%&64.0\%&54.4\%&91.1\%&37.7\%\\
\hline
Adv. Duration +/-&2&91.7\%&1.5\%&2.7\%&0.6\%&0.2\%&2.5\%&7.2\%&10.9\%&2.5\%&82.8\%&\textbf{81.3\%}&71.2\%&38.3\%\\
Adv. Duration +&3&95.2\%&9.4\%&1.6\%&0.3\%&0.2\%&17.5\%&11.6\%&12.5\%&5.9\%&80.2\%&70.1\%&91.1\%&26.8\%\\
20x Duration&4&94.8\%&9.2\%&4.4\%&0.1\%&0.1\%&34.5\%&23.6\%&16.1\%&15.4\%&73.0\%&67.3\%&91.2\%&32.4\%\\
100x Duration&5&95.8\%&7.5\%&5.9\%&3.0\%&0.2\%&40.2\%&29.9\%&16.9\%&21.2\%&\textbf{86.4\%}&80.1\%&91.5\%&46.4\%\\
\hline
Adv. Set 2 +/-&6&93.4\%&2.9\%&3.6\%&0.8\%&0.4\%&0.3\%&0.3\%&0.7\%&0.7\%&64.6\%&55.0\%&88.5\%&11.3\%\\
Adv. Set 2 +&7&94.0\%&5.7\%&3.3\%&1.2\%&0.4\%&1.4\%&0.3\%&1.1\%&0.8\%&82.0\%&69.0\%&\textbf{96.9\%}&23.9\%\\
Adv. Set 2 server +&8&95.2\%&17.4\%&2.5\%&0.7\%&0.4\%&19.0\%&10.3\%&0.7\%&23.3\%&86.3\%&74.5\%&79.2\%&51.4\%\\
Adv. Set 2 client +&9&93.6\%&3.9\%&2.0\%&0.4\%&0.3\%&4.0\%&1.3\%&12.4\%&0.4\%&78.6\%&68.6\%&87.8\%&11.8\%\\
\hline
Adv. Set 3 +/-&10&78.7\%&\textbf{39.6\%}&35.1\%&86.4\%&\textbf{98.8\%}&\textbf{56.0\%}&\textbf{50.5\%}&33.0\%&\textbf{48.9\%}&0.2\%&0.1\%&11.8\%&6.1\%\\
Adv. Set 3 +&11&87.2\%&30.9\%&28.5\%&78.8\%&97.4\%&26.5\%&16.1\%&27.5\%&12.0\%&0.6\%&0.1\%&5.0\%&3.1\%\\
Adv. Set 3 server +&12&\textbf{96.3\%}&32.8\%&22.5\%&52.7\%&69.1\%&41.9\%&32.3\%&16.9\%&33.1\%&26.0\%&7.7\%&0.2\%&\textbf{61.1\%}\\
Adv. Set 3 client +&13&92.2\%&28.6\%&27.9\%&16.5\%&11.7\%&30.9\%&16.3\%&38.3\%&7.1\%&36.5\%&19.7\%&92.6\%&0.4\%\\

\hline

 \end{tabular}
  \caption{Percentage of adversarial flows generated with attack feature set (column) that are misclassified by a model hardened with another attack feature set (row). Same index for attack in column and hardened model in row.}
  \label{tab:hardening-results}
\end{table*}

\textit{Analysis -- attack performance}. Using all features (\textit{Adv. All Features +/-}, column 1) yields by far the best -- although impracticable under our assumptions -- attack; the only model that is robust to this attack is the model that is hardened with this dataset. Feature set 3 attacks (columns 10-13) are themselves successful in attacking models hardened with the other feature sets (rows 1-9). Notice that \textit{Adv. Set 3 server +}, column 12 causes large percentages of misclassified malicious traffic on most detectors except those hardened with datasets using feature set 3; it falls under our assumption that the proxy is only able to increment counters like packets, bytes, and duration; and it only requires deployment at the server side -- making it a good candidate for successful attacks to these models.

\textit{Analysis -- model robustness}. Adversarial samples from feature set 3 (rows 10-13) do not seem adequate for hardening detectors against other adversarial samples. In fact, most attacks are consistently successful against models hardened with any of the feature sets 3 adversarial samples -- except attacks using feature set 3 itself. The detector hardened with all features (row 1) is surprisingly poorly resistant to all other attacks. Feature set 2-hardened models seem to be  robust against both duration and feature set 2 attacks. One reason why feature set 2-hardened models are robust to duration attacks may be because  of the weight of the duration feature in the feature set 2 compared to feature set 3 where there are much more features and the impact of duration changes may be diluted.

\textit{Analysis -- impact on attack performance of only increasing feature values}. Only increasing feature values causes a small yet relatively consistent drop in misclassification of malicious flows. Table \ref{tab:hardening-results} has 3 cases where we can compare the performance of attacks using the same features sets -- each case with one feature set where the FGSM-suggested feature values can be added or subtracted from the original values, and another where they can only be added. These are columns 2 and 3, 6 and 7, and 10 and 11. Models hardened with adversarial flows whose feature values can be larger or smaller than the original flow also consistently perform better than their increase-limited counterparts. One exception is for  feature set 3, whose \textit{Adv. Set 3 +} hardened model fares better than its \textit{Adv. Set 3 +/-} counterpart.

\subsection{Non-adversarial}

While the defender may prepare for an adversarial attack like the one described in the paper and deploy one of the hardened models, it is possible that the attacker chooses not to deploy any adversarial attack -- not altering malware traffic. In this case we test on the original MTA dataset and observe that all hardened models decrease accuracy, precision, and recall, but only slightly, when compared to the original model. From original model values of 95\% accuracy, 92.7\%  precision, and 2.7\% attack success rate (1-recall), no hardened model drops below 91\% accuracy and 86\% precision nor goes above 6\% attack success rate (1-recall).

\section{Iteratively attacking and hardening}
\label{sec:loop}

\begin{figure}[ht]
  \begin{center}
    \leavevmode
    \includegraphics[width=0.4\textwidth]{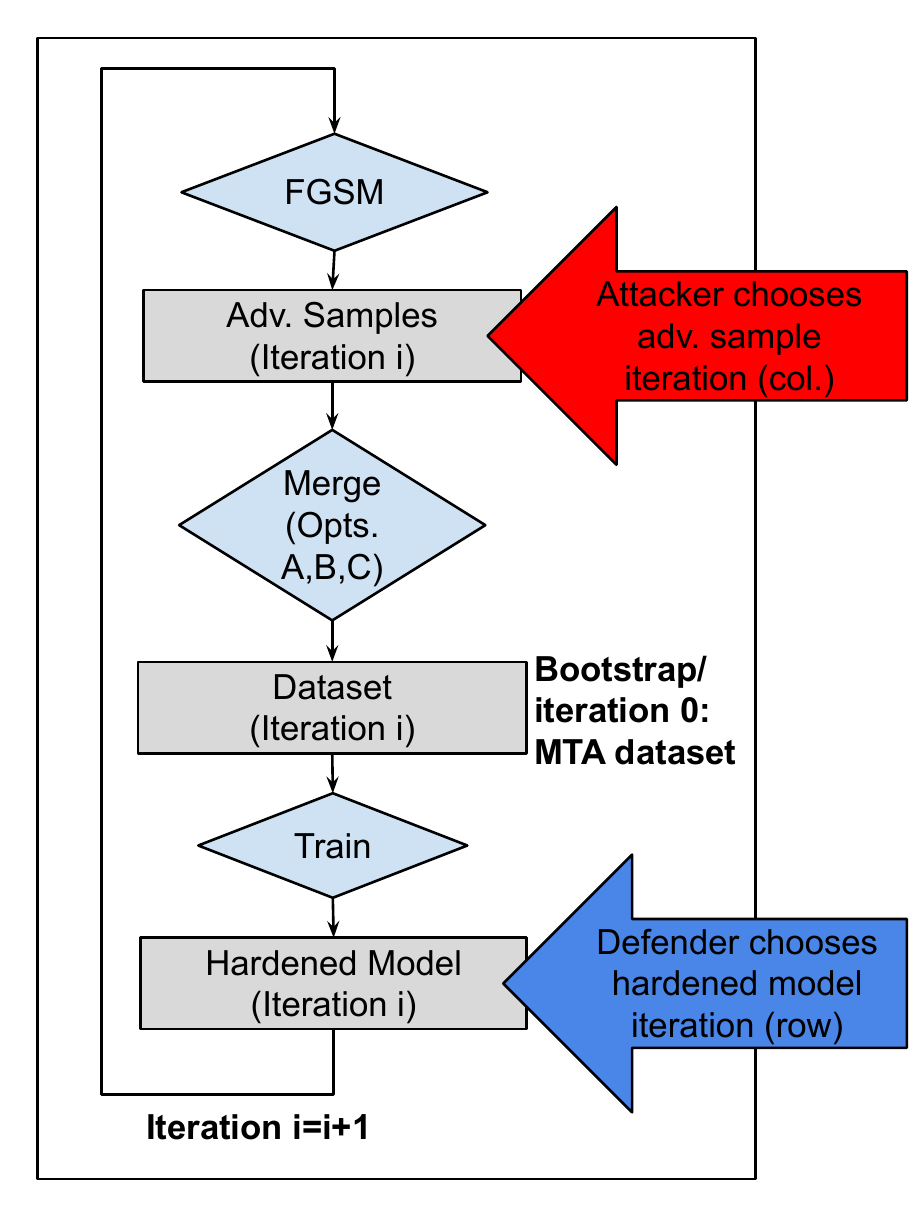}
    \caption{Attack and hardening iterative approach. The bootstrap at iteration 0 starts with the MTA dataset that after training yields the iteration 0 model which is the original non-hardened model.}
    \label{fig:iter_fig6}
  \end{center}
\end{figure}

In this section we try to understand to which extent a model hardened with adversarial samples from one iteration is robust against samples from another iteration. Figure \ref{fig:iter_fig6} illustrates our approach, where defender and attacker use the same iterative approach but where the defender and attacker may independently choose different iterations for hardened model (defender) and adversarial samples (attacker) respectively.

To incrementally attack and harden the detectors we train a detector, at each iteration, with three different options for datasets: A) use original benign flows and the adversarial flows of the previous iteration (original malicious flows if first iteration); B) use original benign and malicious flows together with the adversarial flows of the previous iteration (none if first iteration); C) original benign and malicious flows together with the adversarial flows of all prior iterations (none if first iteration). While training options A and B have fixed training size, option C has increasing training size and may not be sustainable for large number of iterations and large datasets. In the three cases we use all features for the adversarial attack. Table \ref{tab:hardening-loop-results} shows the resilience to adversarial attacks on models hardened in different iterations.

Hardening-loop option C is robust against adversarial attacks at any iteration; this points to C being the best defense, although with growing datasets and iterations this option may be impractical. Option A yields hardened models that are not able to detect the original malicious flows (cf. \textit{MTA} column) -- given that the training dataset for this option at each iteration only includes benign flows and adversarial flows from the previous iteration. Option B is able to detect original malicious flows and has fixed training set size. For option B we observe that, unlike C, no hardened model is robust against the adversarial samples of all iterations, raising a level playing field problem for the defender and attacker. Depending on the iterations that attacker and defender independently choose for their hardened model and adversarial samples, a more successful attack or defense will take place.

We notice that for options B and C and at some iterations the adversarial attack is not successful -- for example option B iteration 3 hardened model only misclassifies 8.9\% of the adversarial flows obtained by directly applying FGSM to its model (row \textit{Adv. 4}); most other models are vulnerable to their adversarial attacks (bold and italic entries in the table). This may be a problem for the attacker since the attack at that iteration is ineffective, and for the defender because the next iteration model hardening does not increase or decrease robustness against this attack. We also notice that all adversarial samples are successful against the original model (cf. \textit{Original} rows), and that  models seem to be yield low misclassified adversarial flows against adversarial attacks not only of the current iteration but also of the next iteration. 

The most important thing to notice here is probably that under this attacking-hardening loop and depending on the strategy for reusing data (A,B,C) there can be a level playing field between attacker and defender where winning is not simply hardening your model as much as you can – or getting the adversarial samples from the most hardened models – if you don’t know how many iterations your opponent decided to do in their attacking-hardening loop. For example in option B the Adv. Iter 5 row performs acceptably for adversarial samples 3, 4, and 5 but not for samples 2 and 1, nor even for the original MTA. This underpins the non-monotonic variation of values in the same rows or columns in table 7. 

\begin{table}[ht]
\small
  \centering
\begin{tabular}{|r||r||r|r|r|r|r|}	
\hline

\textbf{A}&MTA&Adv.1&Adv.2&Adv.3&Adv.4&Adv.5\\
\hline
Original&\textbf{2.7\%}&\textbf{\textit{95.0\%}}&93.6\%&97.0\%&99.8\%&94.9\%\\
Mod. Iter 1&5.7\%&\textbf{0.1\%}&\textbf{\textit{93.1\%}}&23.6\%&0.0\%&93.4\%\\
Mod. Iter 2&99.8\%&0.1\%&\textbf{0.0\%}&\textbf{\textit{100.0\%}}&0.0\%&12.4\%\\
Mod. Iter 3&100.0\%&97.6\%&0.1\%&\textbf{0.1\%}&\textbf{\textit{91.7\%}}&40.4\%\\
Mod. Iter 4&100.0\%&18.0\%&99.9\%&0.0\%&\textbf{0.0\%}&\textbf{\textit{100.0\%}}\\
Mod. Iter 5&100.0\%&71.8\%&35.2\%&100.0\%&0.0\%&\textbf{0.0\%}\\
\hline
\hline
\textbf{B}&MTA&Adv.1&Adv.2&Adv.3&Adv.4&Adv.5\\
\hline
Original&\textbf{2.7\%}&\textbf{\textit{95.0\%}}&92.1\%&100.0\%&98.0\%&99.5\%\\
Mod. Iter 1&5.1\%&\textbf{0.1\%}&\textbf{\textit{90.2\%}}&2.9\%&0.1\%&32.5\%\\
Mod. Iter 2&4.6\%&0.1\%&\textbf{0.2\%}&\textbf{\textit{59.7\%}}&0.1\%&78.6\%\\
Mod. Iter 3&2.9\%&85.4\%&0.1\%&\textbf{0.1\%}&\textbf{\textit{8.9\%}}&32.1\%\\
Mod. Iter 4&4.2\%&25.4\%&59.8\%&0.1\%&\textbf{0.1\%}&\textbf{\textit{32.9\%}}\\
Mod. Iter 5&8.0\%&18.4\%&55.3\%&1.3\%&0.1\%&\textbf{0.2\%}\\
\hline
\hline
\textbf{C}&MTA&Adv.1&Adv.2&Adv.3&Adv.4&Adv.5\\
\hline
Original&\textbf{2.7\%}&\textbf{\textit{95.0\%}}&99.5\%&90.2\%&91.1\%&65.9\%\\
Mod. Iter 1&3.3\%&\textbf{0.2\%}&\textbf{\textit{99.5\%}}&59.5\%&95.4\%&27.9\%\\
Mod. Iter 2&4.8\%&0.1\%&\textbf{0.1\%}&\textbf{\textit{14.3\%}}&0.4\%&10.2\%\\
Mod. Iter 3&8.3\%&0.1\%&0.0\%&\textbf{0.0\%}&\textbf{\textit{0.5\%}}&0.2\%\\
Mod. Iter 4&3.1\%&0.1\%&0.1\%&0.1\%&\textbf{0.0\%}&\textbf{\textit{3.0\%}}\\
Mod. Iter 5&4.9\%&0.1\%&0.1\%&0.0\%&0.1\%&\textbf{0.0\%}\\
\hline
 \end{tabular}
  \caption{Percentage of adversarial flows generated at given iteration (column) that are misclassified by a model hardened with another iteration (row). A, B, C: training option for hardening, details in text. Original model in first row of A, B, and C. }
  \label{tab:hardening-loop-results}
\end{table}

\section{Related Work}
\label{sec:related-work}

Rule- and black-list-based detectors can be complemented by statistical detectors based on traffic characteristics. Statistical detectors can use  packet byte-streams~\cite{wei_wang_malware_2017, marin_deep_2019}, flow features~\cite{anderson_machine_2017}, packet and TLS record count and sizes~\cite{anderson_limitless_2019}, and combinations thereof as input data. As these techniques are further developed we expect different network managers to deploy different solutions or sets of solutions. Adversarial techniques generate samples in the input space of the classifier that increase misclassification and going from classifier input space to actual traffic may be simple -- as in the case of generating a sequence of packets with given size -- or more complex if the input space imposes constraints in the values and if there are limitations to the extent to which the original malicious traffic can be changed. In this paper we focus on flow features; we expect e.g. packet byte-streams to be even harder to generate abiding to constraints but easier to break because the detector has more inputs that can be modified by an attacker. Recent work has focused on generating real packet traffic~\cite{cheng_pac-gan_2019,lin_idsgan_2019} and network flows~\cite{ring_flow-based_2019} to improve datasets used in malware traffic detection using Generative Adversarial Networks. The flow-based approach relies entirely on the GAN to generate adversarial flows, yet it may be impossible to craft such a flow. Furthermore these papers do not consider any limitations on modifying malware to support adversarial attacks without altering the intended malicious behavior of the malware. The packet-based approaches consider some constraints on the structure of the packet but for flow detectors in particular the ability of a sequence of adversarial GAN-generated packets that are misclassified by a packet detector to yield a flow whose statistical features are also misclassified by a flow detector has not been studied. 

\cite{gardiner2016security} surveys command and control traffic detection systems as well as attacks to these systems -- and identify high-level issues including the difficulty of modifying features in C2 traffic and different attack techniques such as poisoning and evasion. In this paper we only consider evasion attacks. Although \cite{gardiner2016security}  looks at different aspects of security for machine learning there are few surveyed examples of actual C2 traffic detection and evasion issues. The following are three examples of recent work that takes on the challenge of attacking C2 traffic detection systems. \cite{rigaki_bringing_2018} describes the use of a GAN to generate adversarial samples that are able to break CTU's Stratosphere IDS. The paper takes an open-source malware and re-codes it so that it supports adversarial samples from the GAN specifying total flow duration and bytes, as well as delay to the next flow. Although the paper is interesting and shows the point of how easy it is to break an IDS, it has significant differences to our work: the detector is not a flow detector but a 3-tuple (victim IP, C2 server IP, C2 server port) detector with features that are not applicable to individual flows, and it makes the assumption that the malware traffic can be modified to support any adversarial feature suggested by the GAN generator, which we do not. More in line with our assumptions about the extent to which C2 traffic can be modified, \cite{apruzzese2018evading} analyzes how flow-based detectors can be evaded by adding a fixed amount to the duration, source and destination byte counts, and total packet counts of the original C2 flow. It uses CTU's 2013 dataset to train a random forest classifier and to show the performance of their attack. \cite{han2020practical} tries to bridge the gap between feature space attacks and their viability in what they call traffic space -- meaning the set of flow feature values that are feasible network-wise and for a given attack. They use a GAN to generate adversarial samples and an optimization procedure to bring the adversary samples close to the misclassification boundary; then they perform traffic mutation to automatically find traffic-space vectors that are close to the adversarial sample. They then harden a detector with the adversarial, traffic-space samples.

Unlike the work we review in this section, in this paper we focus specifically on encrypted malware command and control traffic rather than non-C2 attack traffic or non-encrypted C2 traffic. Moreover, while most related work assesses the performance of the attack on a detector and some work hardens the detector, they do not consider the impact of different feature sets used in the adversarial attack nor of different hardening iterations.

Finally, for non-network related data, a conceptually similar work \cite{laskov2014practical} addresses evasion of PDF malware detection where the authors show that manipulating a small subset of suitable features is effective against a random forest classifier in a black-box setting. They insert dummy content that is ignored by PDF renderers but affects feature values and analyze the impact of different attack scenarios on the detection rate. Analysis of using adversarial examples for hardening led to similar findings to those in section \ref{sec:hardening} with respect to the effectiveness of this kind of defense, but does not explore the attack and hardening loop approach of \ref{sec:detection-and-evasion} -- and does not focus on network traffic.

\section{Conclusions}
\label{sec:conclusions}

In this paper we assessed the performance of evasion attacks and of hardening detector models under the assumption that the malware should maintain its original behavior. For that purpose we use a public source of malware traffic captures to build a labeled dataset from which both defenders and attackers can train and attack C2 encrypted traffic detector models. We show the impact of the practical limitations on a specific attack that we implement on the packet traces, with increasing misclassification of adversarial samples. We then harden the detector with different feature set adversarial samples. Assuming that the detector does not know which feature set the attacker uses and that the attacker does not know which feature set the defender uses to harden the detector, we look for combinations of feasible attacks and models that would either make it impossible for the attacker to win or impossible for the defender to win. We find that an attack that allows adding or subtracting adversarial values from all original features is the best attack -- but is not implementable under our attacker assumptions -- yet does not yield the best hardened models at large. Finally, we try to understand the same issues for different iterations of attacking and hardening a model. We find out that it is feasible to reach a level playing field where for a given number of iterations and a given training set strategy no hardened model is able to detect the attacks of all iterations nor can an attack cause  detectors at all iterations to misclassify a large part of its adversarial samples. 

We intend to implement the proxy-based attacks for feature sets 2 and 3 both from client and server size, test the performance and impact on feature set- and iteration-hardening of other adversarial learning methods, and build and use a more extensive C2 encrypted data set. We also plan to extend this work to other input data -- including packet sizes, TLS records, and packet byte-streams. 

To ease the reproducibility of results we can share parts of our code on-demand -- although most of what is needed is available publicly: the MTA dataset, Tstat, and the Cleverhans adversarial learning framework; the learning models are very simply and easy to code using e.g. Kereas and Tensorflow.

\bibliographystyle{acm}
\bibliography{aisec2020}

\begin{thebibliography}{10}

\bibitem{anderson_limitless_2019}
{\sc Anderson, B., Chi, A., Dunlop, S., and McGrew, D.}
\newblock Limitless {HTTP} in an {HTTPS} {World}: {Inferring} the {Semantics}
  of the {HTTPS} {Protocol} without {Decryption}.
\newblock In {\em Proceedings of the {Ninth} {ACM} {Conference} on {Data} and
  {Application} {Security} and {Privacy}\/} (Richardson, Texas, USA, Mar.
  2019), {CODASPY} '19, Association for Computing Machinery, pp.~267--278.

\bibitem{anderson_machine_2017}
{\sc Anderson, B., and McGrew, D.}
\newblock Machine {Learning} for {Encrypted} {Malware} {Traffic}
  {Classification}: {Accounting} for {Noisy} {Labels} and {Non}-{Stationarity}.
\newblock In {\em Proceedings of the 23rd {ACM} {SIGKDD} {International}
  {Conference} on {Knowledge} {Discovery} and {Data} {Mining}\/} (New York, NY,
  USA, 2017), {KDD} '17, ACM, pp.~1723--1732.

\bibitem{anderson_deciphering_2018}
{\sc Anderson, B., Paul, S., and McGrew, D.}
\newblock Deciphering malware’s use of {TLS} (without decryption).
\newblock {\em Journal of Computer Virology and Hacking Techniques 14}, 3 (Aug.
  2018), 195--211.

\bibitem{apruzzese2018evading}
{\sc Apruzzese, G., and Colajanni, M.}
\newblock Evading botnet detectors based on flows and random forest with
  adversarial samples.
\newblock In {\em 2018 IEEE 17th International Symposium on Network Computing
  and Applications (NCA)\/} (2018), IEEE, pp.~1--8.

\bibitem{cheng_pac-gan_2019}
{\sc Cheng, A.}
\newblock {PAC}-{GAN}: {Packet} {Generation} of {Network} {Traffic} using
  {Generative} {Adversarial} {Networks}.
\newblock In {\em 2019 {IEEE} 10th {Annual} {Information} {Technology},
  {Electronics} and {Mobile} {Communication} {Conference} ({IEMCON})\/} (Oct.
  2019), pp.~0728--0734.
\newblock ISSN: 2644-3163.

\bibitem{demontis2019adversarial}
{\sc Demontis, A., Melis, M., Pintor, M., Jagielski, M., Biggio, B., Oprea, A.,
  Nita-Rotaru, C., and Roli, F.}
\newblock Why do adversarial attacks transfer? explaining transferability of
  evasion and poisoning attacks.
\newblock In {\em 28th $\{$USENIX$\}$ Security Symposium ($\{$USENIX$\}$
  Security 19)\/} (2019), pp.~321--338.

\bibitem{gardiner2016security}
{\sc Gardiner, J., and Nagaraja, S.}
\newblock On the security of machine learning in malware {C\&C} detection: A
  survey.
\newblock {\em ACM Computing Surveys (CSUR) 49}, 3 (2016), 1--39.

\bibitem{goodfellow_explaining_2015}
{\sc Goodfellow, I.~J., Shlens, J., and Szegedy, C.}
\newblock Explaining and {Harnessing} {Adversarial} {Examples}.
\newblock {\em arXiv:1412.6572 [cs, stat]\/} (Mar. 2015).
\newblock arXiv: 1412.6572.

\bibitem{han2020practical}
{\sc Han, D., Wang, Z., Zhong, Y., Chen, W., Yang, J., Lu, S., Shi, X., and
  Yin, X.}
\newblock Practical traffic-space adversarial attacks on learning-based nidss.
\newblock {\em arXiv preprint arXiv:2005.07519\/} (2020).

\bibitem{laskov2014practical}
{\sc Laskov, P., et~al.}
\newblock Practical evasion of a learning-based classifier: A case study.
\newblock In {\em 2014 IEEE symposium on security and privacy\/} (2014), IEEE,
  pp.~197--211.

\bibitem{lin_idsgan_2019}
{\sc Lin, Z., Shi, Y., and Xue, Z.}
\newblock {IDSGAN}: {Generative} {Adversarial} {Networks} for {Attack}
  {Generation} against {Intrusion} {Detection}.
\newblock {\em arXiv:1809.02077 [cs]\/} (June 2019).
\newblock arXiv: 1809.02077.

\bibitem{marin_deep_2019}
{\sc Marín, G., Casas, P., and Capdehourat, G.}
\newblock Deep in the {Dark} - {Deep} {Learning}-{Based} {Malware} {Traffic}
  {Detection} {Without} {Expert} {Knowledge}.
\newblock In {\em 2019 {IEEE} {Security} and {Privacy} {Workshops} ({SPW})\/}
  (May 2019), pp.~36--42.

\bibitem{papernot2018cleverhans}
{\sc Papernot, N., Faghri, F., Carlini, N., Goodfellow, I., Feinman, R.,
  Kurakin, A., Xie, C., Sharma, Y., Brown, T., Roy, A., Matyasko, A., Behzadan,
  V., Hambardzumyan, K., Zhang, Z., Juang, Y.-L., Li, Z., Sheatsley, R., Garg,
  A., Uesato, J., Gierke, W., Dong, Y., Berthelot, D., Hendricks, P., Rauber,
  J., and Long, R.}
\newblock Technical report on the cleverhans v2.1.0 adversarial examples
  library.
\newblock {\em arXiv preprint arXiv:1610.00768\/} (2018).

\bibitem{rigaki_bringing_2018}
{\sc Rigaki, M., and Garcia, S.}
\newblock Bringing a {GAN} to a {Knife}-{Fight}: {Adapting} {Malware}
  {Communication} to {Avoid} {Detection}.
\newblock In {\em 2018 {IEEE} {Security} and {Privacy} {Workshops} ({SPW})\/}
  (May 2018), pp.~70--75.

\bibitem{ring_flow-based_2019}
{\sc Ring, M., Schlör, D., Landes, D., and Hotho, A.}
\newblock Flow-based network traffic generation using {Generative}
  {Adversarial} {Networks}.
\newblock {\em Computers \& Security 82\/} (May 2019), 156--172.

\bibitem{tstat}
{\sc Trevisan, M., Finamore, A., Mellia, M., Munafo, M., and Rossi, D.}
\newblock Traffic {Analysis} with {Off}-the-{Shelf} {Hardware}: {Challenges}
  and {Lessons} {Learned}.
\newblock {\em IEEE Commun. Mag. 55}, 3 (2017), 163--169.

\bibitem{wei_wang_malware_2017}
{\sc Wang, W., Zhu, M., Zeng, X., Ye, X., and Sheng, Y.}
\newblock Malware traffic classification using convolutional neural network for
  representation learning.
\newblock In {\em 2017 {International} {Conference} on {Information}
  {Networking} ({ICOIN})\/} (Jan. 2017), pp.~712--717.

\end{thebibliography}

\section*{Acknowledgements}

This work is financed by the ERDF -- European Regional Development Fund through the Operational Programme for Competitiveness and Internationalisation -- COMPETE 2020 Programme and by National Funds through the Portuguese funding agency, FCT -- \textit{Funda\c{c}\~{a}o para a Ci\^{e}ncia e a Tecnologia} within project POCI-01-0145-FEDER-032454 (PTDC/EEI-TEL/32454/2017).

\end{document}